\documentclass{IEEEtran4PSCC}
\ifCLASSINFOpdf
   \usepackage[pdftex]{graphicx}
\else
   \usepackage[dvips]{graphicx}
\fi
\usepackage[cmex10]{amsmath}

\renewcommand{\baselinestretch}{0.97}

\makeatletter
\let\old@ps@headings\ps@headings
\let\old@ps@IEEEtitlepagestyle\ps@IEEEtitlepagestyle
\def\psccfooter#1{%
    \def\ps@headings{%
        \old@ps@headings%
        \def\@oddfoot{\strut\hfill#1\hfill\strut}%
        \def\@evenfoot{\strut\hfill#1\hfill\strut}%
    }%
    \def\ps@IEEEtitlepagestyle{%
        \old@ps@IEEEtitlepagestyle%
        \def\@oddfoot{\strut\hfill#1\hfill\strut}%
        \def\@evenfoot{\strut\hfill#1\hfill\strut}%
    }%
    \ps@headings%
}
\makeatother

\usepackage{amsfonts}
\usepackage{caption} 
\usepackage{subcaption} 
\usepackage{graphicx}

\begin{document}
%
\title{Flexibility from Networks of Data Centers: \\ A Market Clearing Formulation with Virtual Links}



 \author{\IEEEauthorblockN{Weiqi Zhang\IEEEauthorrefmark{1},
 Line A. Roald\IEEEauthorrefmark{1},
 Andrew A. Chien\IEEEauthorrefmark{2}\IEEEauthorrefmark{3}, 
 John R. Birge\IEEEauthorrefmark{2} and
 Victor M. Zavala\IEEEauthorrefmark{1}}
 \IEEEauthorblockA{\IEEEauthorrefmark{1}University of Wisconsin-Madison, Madison, WI, USA, \IEEEauthorrefmark{2}University  of Chicago, Chicago, IL, USA, \\
 \IEEEauthorrefmark{3}Argonne National Laboratory, Lemont, IL, USA}
}\date{June 2019}

\maketitle

\begin{abstract}

Data centers owned and operated by large companies have a high power consumption and this is expected to increase in the future. However, the ability to shift computing loads geographically and in time can
provide flexibility to the power grid. We introduce the concept of virtual links to capture space-time load flexibility provided by geographically-distributed data centers in market clearing procedures. 
We show that the virtual link abstraction fits well into existing market clearing frameworks and can help analyze and establish market design properties. This is demonstrated using illustrative case studies.
\end{abstract}


\section{Introduction}



Information and computing technologies are the fastest growing uses of electric power (accounting for 8\% in 2016 and projected at 13\% by 2027), implying that the {\em dynamics and spatial layouts} of computing loads will increasingly affect power grid operations.  
At the same time, computing infrastructures are undergoing major structural changes. 
First, enterprise and government computing functions are being {\em consolidated into fewer and larger scale facilities}, to support efficient and reliable processing of time-sensitive workloads like search services as well as large-scale workloads that run more effectively on large shared computing facilities \cite{barroso_holzle_ranganathan_2018}. This centralization benefits from economies of scale, such as increased utilization and energy efficiency, but concentrates power loads in
a smaller number of locations.  According to the Natural Resources Defense Council \cite{nrdc_2014}, U.S. data centers in 2013 consumed 90 TWh (the output of 34 power plants with capacities of 500 MW) which is projected to grow to 140 TWh by 2020 (an increase of 55\%). To give some perspective on the magnitude of data center loads, we note that large data centers are in the order of 100 MW, with potentially several nearby locations. 
It has been recently reported that training a single machine learning (ML) model can consume as much as 500 MWh and ML models are documented as growing in size at as rates as high as 10x per year \cite{amodei_2019}. 

Second, emergence of ``hyperscalars'' (e.g., Amazon, Google,
Facebook, Microsoft, Alibaba, Tencent) that provide consumer internet and cloud computing services to billions of users has lead to the emergence of networks of data centers. These networks are managed and controlled {\em collectively} (typically via network operation centers (NOCs)). The consumption of these collections are already large, for example, Google's collection consumed 26 TWh in 2017, and is growing fast \cite{dreyfuss_2018}. 
Across these networks, computing tasks can be modulated at multiple timescales (from milliseconds to hours) and can be shifted geographically over long distances (within and beyond the boundaries of independent system operators (ISOs) managing the power grid). This enables {\em space-time shifting flexibility} of the associated power loads. 

Previous research has looked into various ways to harness space-time flexibility from data centers via demand response programs. Data centers are able to provide temporal flexibility (i.e., as a storage device) \cite{liu_liu_low_wierman_2014}. In low-to-medium power density data centers, this can be achieved using pre-cooling strategies 
\cite{lukawski_tester_moore_krol_anderson_2019}. In more efficient data centers, strategies to shift loads include server idling, load migration, shutdown and idling of servers and storage clusters, and cooling relative to load reduction \cite{ghatikar_ganti_matson_piette_2012}, \cite{li_bao_li_2015}. 
Research has also focused on using load shifting and demand response in data centers to enable incorporation of renewable energy sources.
Strategies include 
power cappings to match demand and renewable energy supply \cite{gmach_rolia_wang_2010}, shifting loads to reduce peak demand in the face of uncertainty \cite{liu_wierman_chen_razon_chen_2013}, and geographical redistribution to facilitate renewable adoption \cite{wang_ye_2016}, \cite{liu_lin_werman_low_andrew_2011}.
The effectiveness of shifting computing loads to minimize cost has been demonstrated in
\cite{rao_liu_xie_liu_2010, li_bao_li_2015, rao_liu_ilic_liu_2012}, considering shifts between multiple electricity markets \cite{rao_liu_xie_liu_2010} and cooperation between data centers \cite{rao_liu_ilic_liu_2012}.

While the potential of harnessing flexibility of data centers in power grid operations is promising, gathering such flexibility can be challenging due to the complex nature of data center workloads \cite{dcdr_survey}. Therefore, it is necessary to analyze how electricity market designs can influence and incentivize provision of flexibility.
The use of online auctioning models to study incentives needed to harness spatial and temporal flexibility from data centers has been studied in \cite{edr_temporal_online_auction}, \cite{edr_colo_online_auction}. A Nash bargaining formulation is employed to analyze interactions between data centers and load serving entities \cite{dcdr_nash_bargaining} and between data centers and tenants \cite{colo_dcdr_nash}.

In this work, we propose market clearing models that capture space-time flexibility provided by load shifting and migration in data centers. Shifting can be achieved by a company owning a set of geographically distributed data centers, which allows management of where and when to satisfy the loads. Specifically, the company can choose to shift the workload to servers at another location or to delay the workload to later times.  We note that previous work has studied models and mechanisms for the provision of load flexibility from a data center perspective. Limited studies have analyzed load shifting mechanisms and models from a systems-wide perspective (ISO perspective). For instance, recent work has revealed that shiftable loads can help absorb stranded power and control power flows in the network  \cite{kim_yang_zavala_chien_2017,yang_chien_2016}.  Here, we study the load shifting problem from a coordinated market clearing perspective and present formulations that accommodate space-time shifting flexibility. We take the perspective of an ISO, which seeks to clear the market by using demand and supply bidding information in a coordinated manner. Recent examples  \cite{gribik_chatterjee_navid_zhang_2011,zavala_kim_anitescu_birge_2017,pritchard_zakeri_philpott_2010,carrion_arroyo_2006,bouffard2005market} seek to analyze how changes in market design (i.e., in the bidding process and clearing formulation) can influence dispatch and price behavior and can benefit ISO operations and market participants.  

The main contributions of this paper are as follows: We introduce the notion of {\em virtual links}, which are (non-physical) pathways that shift loads in space (by reallocating computing loads to other geographical location) and time (by delaying a computing load). The network of virtual links acts as an additional infrastructure layer on top of the existing transmission grid. We show that this paradigm is intuitive and compatible with existing market clearing procedures (in which flexibility is provided by a physical transmission network and generator dynamics). The proposed framework also reveals which information should be provided by data centers in the bidding process and provides insights into conditions that will incentivize data centers to provide flexibility. We also show that virtual links provide a convenient framework to establish pricing and social welfare properties and to analyze complex space-time behavior. Specifically, by means of case studies, we demonstrate that data center flexibility leads to higher social welfare and to higher total loads delivered to data centers. Moreover, we also demonstrate that virtual links lead to local marginal prices (LMPs) that tend to become more spatially and temporally homogeneous because virtual links help relieve transmission network congestion and generator ramping limits.

The paper first presents the market clearing setting in Section \ref{sec:marketclearingsetting}. We start from the traditional setting without any virtual links, and then introduce both spatial and temporal shifts. Section \ref{sec:casestudies} then demonstrates the benefits on different case studies, while Section \ref{sec:conclusion} summarizes and concludes.


\section{Market Clearing Setting}
\label{sec:marketclearingsetting}
The market setting presented here builds on formulations presented in \cite{pritchard_zakeri_philpott_2010,zavala_kim_anitescu_birge_2017}, but is adapted to capture space-time load shifting flexibility. This first part of the section discusses a baseline setting where no load shifts are captured in the market clearing procedure (shifts are introduced in Section \ref{sec:clearing}). 

We begin by considering a time-independent network (of the physical electric grid) with a set of geographical nodes $\mathcal{N} = \{n_0,n_1,...,n_N\}$, suppliers $\mathcal{G}$, loads (data centers) $\mathcal{D}$, and physical transmission lines $\mathcal{L}$.  We also consider a set of entities or stakeholders that own (or operate) loads $\mathcal{E}$. Each generator $i \in \mathcal{G}$ is connected to node $n(i) \in \mathcal{N}$, has a total flow $p_i \in \mathbb{R}_+$, maximum supply capacity $\bar{p}_i \in \mathbb{R}_+$, and bidding cost $\alpha^s_i \in \mathbb{R}_+$. For convenience, we define the set of generators connected to node $n$ as $\mathcal{G}_{n} = \{i \in \mathcal{G} | n(i)=n\}$.   

Each load $j \in \mathcal{D}$ is connected to node $n(j) \in \mathcal{N}$, has a served load $d_{j} \in \mathbb{R}_+$, requested load  $\bar{d}_{j} \in \mathbb{R}_+$, and bidding value $\alpha^d_{j} \in \mathbb{R}_+$. The served loads refer to the outcome of the market clearing. We define the set of loads connected to node $n$ as $\mathcal{D}_{n} \subseteq \mathcal{D}$.  
Each load has an associated owning/operating entity $e(j)\in \mathcal{E}$ and we define the set of of loads owned by entity $e$ as $\mathcal{D}_e\subseteq \mathcal{D}$. Note that the notion of set of loads can be generalized to include both data center and non-datacenter loads. Each link $l \in \mathcal{L}$ has an associated receiving (end) node $\textrm{rec}(l)$, sending (source) node $\textrm{snd}(l)$, physical (bidirectional) power flow $f_{l} \in \mathbb{R}$, maximum capacity $\bar{f}_{l}$, and bidding cost $\alpha_{l}^f$.  We use $\mathcal{L}_n^{in}$ to denote the set of lines with flow that enter node $n$ and $\mathcal{L}_n^{out}$ to denote the set of lines with flow leaving node $n$. We define the market clearing prices (locational marginal prices-LMPs) at node $n$ as $\pi_{n}\in\mathbb{R}$.

Based on the setting described, a basic market clearing formulation (solved by the ISO) takes the form:
\begin{subequations}
\begin{align}
& \underset{(d,p,f) \in \mathcal{C}}{\text{max}} && \phi:=\sum_{j\in\mathcal{D}} \alpha_{j}^d d_{j}-\sum_{i \in \mathcal{G}} \alpha_{i}^p p_{i} -\sum_{l\in \mathcal{L}}\alpha_l^f|f_l| \label{opt:p1_obj} \\ 
& {\text{s.t.}} && \sum_{l \in \mathcal{L}_n^{in}} f_{l}  + \sum_{i \in \mathcal{G}_n} p_{i} =  \sum_{l \in \mathcal{L}_n^{out}} f_{l}+ \sum_{j \in \mathcal{D}_n} d_{j}, n \in \mathcal{N} \label{opt:p1_balance} \\ 
& && f_l=B_l(\theta_{\textrm{snd}(l)}-\theta_{\textrm{rec}(l)}), \quad l \in \mathcal{L} \label{opt:p1_dc_flow}
\end{align}
\end{subequations}
where $\mathcal{C} = \{(d,p,f) | 0 \leq p_{i} \leq \bar{p}_{i}$,  $0 \leq d_{j} \leq \bar{d}_{j}, |f_{l}| \leq \bar{f}_{l}\}$ denotes capacity constraints of generators, loads, and flows. The objective function \eqref{opt:p1_obj} is the social welfare, which seeks to maximize value of the total load served while minimizing operating costs associated with generation and physical transmission. Constraints \eqref{opt:p1_balance} enforce power conservation at each node. Equations \eqref{opt:p1_dc_flow} denote DC flow constraints, where $\theta_n$ is the voltage angle at node $n$ and $B_l$ is the susceptance of line $l$. Note that the proposed model is different from standard DC-OPF formulations in that it explicitly considers transmission cost terms in the social welfare. This is done in order to highlight how load shifting can be remunerated by the market. 

A solution of the clearing problem can also be found by solving the Lagrangian dual problem,
\begin{align}
\max_\pi \; \min_{(d,p,f) \in \mathcal{F}} \mathcal{L}(d,p,f)
\label{eq:maxmin}
\end{align}
where the partial Lagrange function is given by
\begin{align}
&\mathcal{L}(d,p,f)\nonumber
=\sum_{i \in \mathcal{G}} \alpha_{i}^p p_{i} +\sum_{l\in \mathcal{L}}\alpha_l^f|f_l|-\sum_{j\in\mathcal{D}} \alpha_{j}^d d_{j}\nonumber\\
&\; -\sum_{n\in\mathcal{N}}\pi_n\left(\sum_{l \in \mathcal{L}_n^{in}} f_{l}  + \sum_{i \in \mathcal{G}_n} p_{i} -  \sum_{l \in \mathcal{L}_n^{out}} f_{l}- \sum_{j \in \mathcal{D}_n} d_{j}\right)\nonumber\\
&=\sum_{j\in\mathcal{D}}\phi_j+\sum_{i\in\mathcal{G}}\phi_i+\sum_{\ell\in\mathcal{L}}\phi_\ell.
\label{eq:lagrangian}
\end{align}
Here, the feasible set $\mathcal{F}$ includes the set $\mathcal{C}$ and the set of feasible injections induced by the DC constraints \cite{pritchard_zakeri_philpott_2010}. This indicates that the clearing problem finds dual variables (LMPs) $\pi_n$ for the balance constraints that maximize the profit functions for generators $\phi_i:=(\pi_{n(i)}-\alpha_i^s)p_i$,  loads $\phi_j:=(\alpha_j^d-\pi_{n(j)})d_j$, and transmission  lines $\phi_l:=(|\Delta \pi_{l}|-\alpha_l^f)|f_l|$ (where $\Delta \pi_l:=\pi_{\textrm{rec}(l)}-\pi_{\textrm{snd}(l)}$ are the nodal price differences) \cite{sampat2019coordinated}. Intuitively, these profit functions evaluate how much welfare each stakeholder gains (or loses if negative) from the market clearing outcome (in addition to their individual bids). If the set $\mathcal{F}$ accepts zero as a feasible clearing solution (all clearing quantities are zero), one can establish that the profits of all players is non-zero \cite{sampat2019coordinated,pritchard_zakeri_philpott_2010,zavala_kim_anitescu_birge_2017}. From the above formulation we can also see that a bid for a transmission line can be interpreted as an operating cost. This observation will be relevant when exploring properties of virtual load shifting. 
%

\subsection{Market Clearing with Spatial Shifts}\label{sec:clearing}

We now introduce the concept of {\em virtual links} to capture {\em spatial load shifting} flexibility provided by data centers. Consider that an entity $e$ operating a set of data centers $\mathcal{D}_e$ at nodes $\mathcal{N}_e$ offers flexibility to the ISO by allowing to shift part of a load $d_j, j\in\mathcal{D}_e$ from the base node $n(j)$ (which we refer to as the {\em hub node}) to a set of alternate nodes $\mathcal{N}_e$. As mentioned in the introduction, this event can be by methods such as computing load migration, where part of the computing loads at one cluster is migrated to another cluster.
The shifted computing load is received at the source cluster, but executed at the destination cluster.
This hence shifts the demand of electricity from one cluster to the other.
Since the load shifting requires at least two data centers at different locations, this type of flexibility can be offered by entities that own and operate multiple data centers at different geographical locations (e.g., a computational task and associated power load can be executed at different locations). 
We assume that each data center consumes the same amount of energy for a given computing task.

To capture load shifting flexibility in the market clearing, we use \emph{virtual links} as non-physical pathways to quantify this kind of load shifting capability between pairs of clusters located at different nodes, in units of electric power consumed by the shifted jobs. 
In this way, we can think of data centers owned by on entity as an extra layer of transmission network that does not have to adhere to Kirchoff's laws. We will show later that the alternative way of transmission provided by data center flexibility helps clear more loads and reduce electricity price volatility by overcoming physical constraints.

We define a set of {virtual links} for the load requesting entity $e\in\mathcal{E}_d$ as $\mathcal{V}_e$. Each link $v\in\mathcal{V}_e$ has capacity to send a portion of the served load $d_j$ (denoted as $\delta_v\in \mathbb{R}$) from node $n(j)$ to node $n(j')$ with $n(j),n(j')\in\mathcal{N}_e$ (i.e., $\textrm{snd}(v)=n(j)$ and $\textrm{rec}(v)=n(j')$).  To avoid degeneracy, we assume that load cannot be sent and received at the same node (no virtual link $v$ exists with $\textrm{snd}(v)=n(j)$ and $\textrm{rec}(v)=n(j)$). A virtual link provides a {\em non-physical} pathway for the shifted load $\delta_v$ (in the sense that the link does not carry power). As a computing task is shifted to a data center at another node, its associated load is {\em physically} cleared at the destination node. Thus, the total load that is (physically) absorbed at node $n$ is given by:
\begin{align*}
\hat{d}_{n}=\sum_{j\in\mathcal{D}_{n}}d_i+\sum_{v\in\mathcal{V}_{n}^{in}}\delta_v-\sum_{v\in\mathcal{V}_{n}^{out}}\delta_v
\end{align*}
and the total load absorbed at node $n(j)$ needs to satisfy the constraint $0 \leq \hat{d}_{n(j)}\leq d^{max}_n$, where $d^{max}_n$ is the capacity of the data center at node $n$. Here, $\mathcal{V}_n^{in},\mathcal{V}_n^{out}$ are the sets of virtual links that enter and leave node $n$, respectively. If $\delta_v=0$ for all virtual flows $v$ entering and leaving the node $n(j)$, the original load $d_j$ is served at the hub node $n(j)$ and thus $\hat{d}_{n(j)}=\sum_{j\in\mathcal{D}_n}d_j$. On the other hand, if a portion of the load is shifted then this will be physically absorbed at another node. The power conservation equation is thus modified as:
\begin{align*}
&\sum_{l \in \mathcal{L}_n^{in}} f_{l}  +  \sum_{i \in \mathcal{G}_n} p_{i} =  \sum_{l \in \mathcal{L}_n^{out}} f_{l}+\hat{d}_n\\
&\qquad\qquad=  \sum_{l \in \mathcal{L}_n^{out}} f_{l}+\sum_{j \in \mathcal{D}_n}d_j+\sum_{v\in\mathcal{V}_{n}^{in}}\delta_v-\sum_{v\in\mathcal{V}_{n}^{out}}\delta_v.
\end{align*}

Virtual links allow the ISO to shift loads between nodes to overcome physical constraints associated with generation and  transmission. Moreover, we will see that virtual links provide alternative pathways that can be exploited to increase economic efficiency and to mitigate price volatility.  We also observe that adding virtual links is {\em mathematically equivalent} to the introduction of transmission lines with the exception that the flows do not have to adhere to Kirchoff's laws. Note also that, by definition, if an entity $e$ does not own/operate multiple loads (i.e., the sets $\mathcal{D}_e$ and $\mathcal{N}_e$ are singletons) then it cannot offer spatial shifting flexibility. Furthermore, entities that choose to shift power across the network may incur a cost due to increased latency. This provides motivation for thinking of data centers owned by one entity as a network with edges represented by virtual links. Based on these observations it does make sense to assume that virtual links are offered as products in the market clearing framework. Accordingly, we consider the formulation: 
\begin{equation*}
\begin{aligned}
& \underset{(d,p,f,\delta) \in \mathcal{C}}{\text{max}} & & \sum_{j\in\mathcal{D}} \alpha_{j}^d d_{j}-\sum_{i \in \mathcal{G}} \alpha_{i}^p p_{i} - \sum_{l\in \mathcal{L}}\alpha_l^f|f_l| -\sum_{v\in \mathcal{V}}\alpha_v^\delta|\delta_v|\\
& {\text{s.t.}} & &  \sum_{l \in \mathcal{L}	_n^{in}} f_{l}  +  \sum_{i \in \mathcal{G}_n} p_{i} + \sum_{v \in \mathcal{V}_n^{out}} \delta_{v} \\
&&& \qquad=  \sum_{l \in \mathcal{L}_n^{out}} f_{l}+\sum_{j \in \mathcal{D}}d_j+\sum_{v \in \mathcal{V}_n^{in}} \delta_{v}, \; n\in\mathcal{N} \\
&&& f_l=B_l(\theta_{\textrm{snd}(l)}-\theta_{\textrm{rec}(l)}), \quad l \in \mathcal{L}
\end{aligned}
\end{equation*}
where $\mathcal{C} = \{(d,p,f,\delta) |0 \leq p_{i} \leq \bar{p}_{i}, 0 \leq d_{j} \leq \bar{d}_{j}, 0 \leq \hat{d}_{n} \leq d^{max}_n, |f_{l}| \leq \bar{f}_{l}\}$. The formulation can also impose maximum allowable shifts between nodes of the form $\underline{\delta}_l\leq \delta_l\leq \bar{\delta}_l$. These constraints can help capture technical limitations of data centers or asymmetries in allowable shifts (e.g., virtual flows can only move in one direction).

One can directly use duality concepts (similar to eq. \eqref{eq:maxmin}-\eqref{eq:lagrangian}) to show that the above formulation maximizes the profit functions for generators,  loads, and transmission links (as in a standard clearing formulation) and also for virtual links  $\phi_v:=(|\Delta \pi_{v}|-\alpha_v^\delta)|\delta_v|$, where $\Delta \pi_v:=\pi_{\textrm{rec}(v)}-\pi_{\textrm{snd}(v)}$, and $\alpha_v^\delta$ is the cost associated with shifting a load. We thus have that virtual shifting capacity provides an additional revenue stream to market participants.  This also offers mechanism to hedge against bidding risk because offering the opportunity to shift loads will more likely result in more load being served (i.e., the ISO can reconfigure loads to maximize total load served), compared to the case in which an entity requires load at different nodes but does not offer the opportunity to reconfigure them. This behavior is illustrated in Figure \ref{fig:3bus}.

\subsection{Market Clearing with Temporal Shifts}

Although the concept of virtual links naturally arises from the way spatial load shifting is achieved, we can generalize this concept to represent temporal flexibility. Through methods like idling servers, data centers are able to pause the computing jobs when a demand response event starts and restart them afterwards \cite{ghatikar_ganti_matson_piette_2012}. Therefore, we can use a virtual link that branches out from one time point to another in the future to represent the workload delay {\em within} a data center. Similar to the case of spatial flexibility, the virtual link quantifies how much workload is delayed in units of energy consumed, but with a different assumption that no load is dropped due to time delays. We use a cost term in the objective function to capture decrease in quality of service caused by workload delay.

We now show how virtual flows can be used capture {\em temporal load shifting} flexibility.  For simplicity, we consider a single spatial location and define the lexicographic set $\mathcal{T}:=\{t_1,...,t_{T}\}$ to represent a sequence of locations (nodes) in time. The interpretation of time points as nodes will become relevant when capturing space-time behavior. Now consider that the entity $e$ requests loads $d_j,\,j\in\mathcal{D}_e$ to be delivered at times $t_j\in \mathcal{T}$ and thus these nodes are interpreted as the hub nodes. If the loads cannot be met at the hub nodes, the data center offers flexibility to shift a fraction of these loads to other available times, which is captured in the node set $\mathcal{T}_e\subseteq \mathcal{T}$. As before, this is done by using virtual links $\mathcal{V}_e$ that connect the hub nodes to the set of available nodes $\mathcal{T}_e$.  This behavior is illustrated in Figure \ref{fig:5times}.

We observe that virtual flows can also be used to capture storage dynamics. To illustrate this, we assume that the firm requests loads $d_t$ at nodes $t\in\mathcal{T}$ and define the set of virtual links $\mathcal{V}_e:=\{v_1,...,v_{T-1}\}$ with element $v_l$ constructed such that $\textrm{snd}(v_l)=t_l$ and $\textrm{rec}(v_l)=t_{l+1}$ for $l=0,...,T-1$. We note that load balances at the time nodes can be written as:
\begin{align*}
\delta_{t}&=\delta_{t-1}+u_{t},\; t=t_1,...,t_{T}
\end{align*}
with $\delta_{t_1}=u_{t_1}$, $\delta_{T}=0$ and $ u_{t}:=d_t-\hat{d}_{t}$. In other words, the virtual flows $\delta_t$ act as storage that carries over unmet demand $u_t$ over future times. Imposing constraints on virtual flows allow us to control the load carry-over and the bidding cost for the virtual flows can be used to capture the fact that a shift might become increasingly expensive as one moves forward in time (e.g., delaying a computation load becomes increasingly expensive as the delay increases). The storage interpretation assumes that carry over only occurs forward in time but we note that virtual flows can be used for arbitrary shifts in time to allow for strategic reallocation during the clearing procedure.

\subsection{Market Clearing with Spatio-Temporal Shifts}

The virtual flows in time and space can be combined to enable a straightforward generalization to capture spatio-temporal shifting flexibility that facilitates analysis and interpretation of clearing formulations.  We define a set of spatial nodes $\mathcal{N}$ and a set of temporal nodes as $\mathcal{T}$. In a standard clearing formulation, the supply, load, and transmission flows at time $t\in\mathcal{T}$ are defined as $p_{i,t}$, $d_{j,t}$, and $f_{l,t}$.  In a market setting that captures space-time flexibility, an entity $e$ requests loads $d_{j,t}$ at hub nodes $(n(j),t)\in\mathcal{N}\times \mathcal{T}$ but also offers the possibility to shift the loads to a set of space-time nodes $\mathcal{N}_e\times \mathcal{T}$. We define the set of virtual links $\mathcal{V}_e$, where each link $v$ sends a virtual flow $\delta_v\in \mathbb{R}$ (fraction of the load $d_{j,t}$) from the space-time hub node $(n(j),t)$ to another space-time node $(n(j'),t')$; in other words, we have that $\textrm{snd}(v)=(n(j),t)$ and $\textrm{rec}(v)=(n(j'),t')$.  As before, to avoid degeneracy, we assume that load cannot be sent and received at the same space-time node (no virtual link $v$ exists with $\textrm{snd}(v)=(n(j),t)$ and $\textrm{rec}(v)=(n(j),t)$). The total load served at space-time location $(n,t)$ is given by:
\begin{align*}
\hat{d}_{n,t}=\sum_{j\in\mathcal{D}_{n}}d_{j,t}+\sum_{v\in\mathcal{V}_{n,t}^{in}}\delta_v-\sum_{v\in\mathcal{V}_{n,t}^{out}}\delta_v
\end{align*}
The market clearing formulation takes the form:
\begin{equation*}
\begin{aligned}
& \underset{(d,p,f,\delta)\in\mathcal{C}}{\text{max}} & & \sum_{t\in \mathcal{T}}\sum_{j\in\mathcal{D}} \alpha_{j,t}^d d_{j,t}-\sum_{t\in \mathcal{T}}\sum_{i \in \mathcal{G}} \alpha_{i,t}^p p_{i,t}\\
&&&\qquad\qquad\qquad - \sum_{t\in \mathcal{T}}\sum_{l\in \mathcal{L}}\alpha_{l,t}^f|f_{l,t}| -\sum_{v\in \mathcal{V}}\alpha_v^\delta|\delta_v|\\
& {\text{s.t.}} & &  \hspace{-3em} \sum_{l \in \mathcal{L}_n^{in}} f_{l,t}  +  \sum_{i \in \mathcal{G}_n} p_{i,t} + \sum_{v \in \mathcal{V}_{n,t}^{out}} \delta_{v} \\
&&& \hspace{-3em}~~=\!\sum_{l \in \mathcal{L}_n^{out}} f_{l,t} +\!\sum_{j\in \mathcal{D}_{n}}d_{j,t}+\!\sum_{v \in \mathcal{V}_{n,t}^{in}} \delta_{v}, ~~
(n,t)\in\mathcal{N}\times \mathcal{T} \\
&&& \hspace{-3em}f_{l,t}=B_l(\theta_{\textrm{snd}(l),t}-\theta_{\textrm{rec}(l),t}), \qquad\qquad \! (l,t) \in \mathcal{L} \times \mathcal{T}
\end{aligned}
\end{equation*}
where $\mathcal{C} = \{(d,p,f,\delta) ~|~ 0 \leq p_{i,t} \leq \bar{p}_{i,t}, 0 \leq d_{j,t} \leq \bar{d}_{j,t}, 0 \leq \hat{d}_{n,t} \leq d^{max}_{n,t}, |f_{l,t}| \leq \bar{f}_{l,t}, |p_{i,t+1}-p_{i,t}|\leq \bar{\Delta p}_i\}$. The latter constraint in $\mathcal{C}$ captures the ramping constraints for generators, which may create time congestion (analogous to transmission congestion in the spatial domain). We will see that temporal virtual shifts allow us to alleviate congestion generated from ramping constraints while spatial shifts alleviate congestion generated from transmission constraints. 

\section{Case Studies}
\label{sec:casestudies}

In this section we demonstrate various benefits of using virtual links. We present small case studies to illustrate the key results and a large study to show that the results are scalable. 

\begin{table*}[h!t]
\centering
\caption{\small Results for three-bus network with temporal shifting flexibility}
\begin{tabular}{|c|c c|c c c c c c|} 
 \hline
 Scenario & $\bar{\delta}$ (MWh) & $\alpha^\delta_v$ (\$/MWh) & $\phi$ (\$) & $\pi$ (\$/MWh) & $d$ (MWh) & $p$ (MWh) & $f$ (MWh) & $\delta$ (MWh)\\ 
 \hline
 1 & [0,0,0] & 3 & 2920 & [10,30,18] & [40,42.5,40] & [42.5,30,50] & [5,-2.5,-7.5] & [0,0,0] \\ 
 2 & [5,0,0] & 3 & 2980 & [10,20,13] & [40,45,40] & [47.5,27.5,45] & [5,-2.5,-7.5] & [-5,0,0] \\
 3 & [15,0,0] & 3 & 2997.5 & [17,20,16.5] & [40,45,40] & [50,25,50] & [5,-2.5,-7.5] & [-7.5,0,0] \\
 4 & [15,0,15] & 3 & 3000 & [17,20,17] & [40,45,40] & [50,25,50] & [5,0,-5] & [-5,0,5] \\
 5 & [100,100,100] & 3 & 3000 & [17,20,17] & [40,45,40] & [50,25,50] & [5,0,-5] & [-5,0,5] \\
 6 & [15,0,15] & 1 & 3030 & [19,20,19] & [40,45,40] & [50,25,50] & [0,0,0] & [-10,0,10] \\
 7 & [15,0,15] & 0 & 3050 & [20,20,20] & [40,45,40] & [50,25,50] & [0,0,0] & [-10,0,10] \\
 \hline
\end{tabular}
\label{table:3bus_results}
\end{table*}

\subsection{Spatial Shifts in a 3-Bus Network}

We consider a network with three spatial nodes and one time node (i.e., we assume there is no temporal virtual shift). Each node contains one supply and one data center and each pair of nodes contains one physical transmission line and one spatial virtual link. We assume that a single entity operates the three data centers. The generation capacities are set to $\bar{p}=\{50,30,50\}$, load capacities to $\bar{d}=\{40,45,40\}$, transmission capacity to $\bar{f} = \{5,10,10\}$, generation bidding costs to $\alpha^g=\{10,20,10\}$, load bidding prices to $\alpha^d=\{40,30,40\}$, and transmission costs for the links to $\alpha^f=\{2,2,2\}$. For each transmission line $l$, we set $B_l = 0.5$. The system is designed in a way that nodes $n_1$ and $n_3$ have excess in supply while $n_2$ has an excess in load. The system has three virtual links $\mathcal{V}=\{(1,2),(1,3),(2,3)\}$ available to shift load. We solve the market clearing problem under six different scenarios that account for increasing levels of spatial flexibility. 

\begin{figure}[!ht]
	\centering
	\vspace{-10pt}
	\includegraphics[width=3.3in, trim={1cm 3cm 0 2cm},clip]{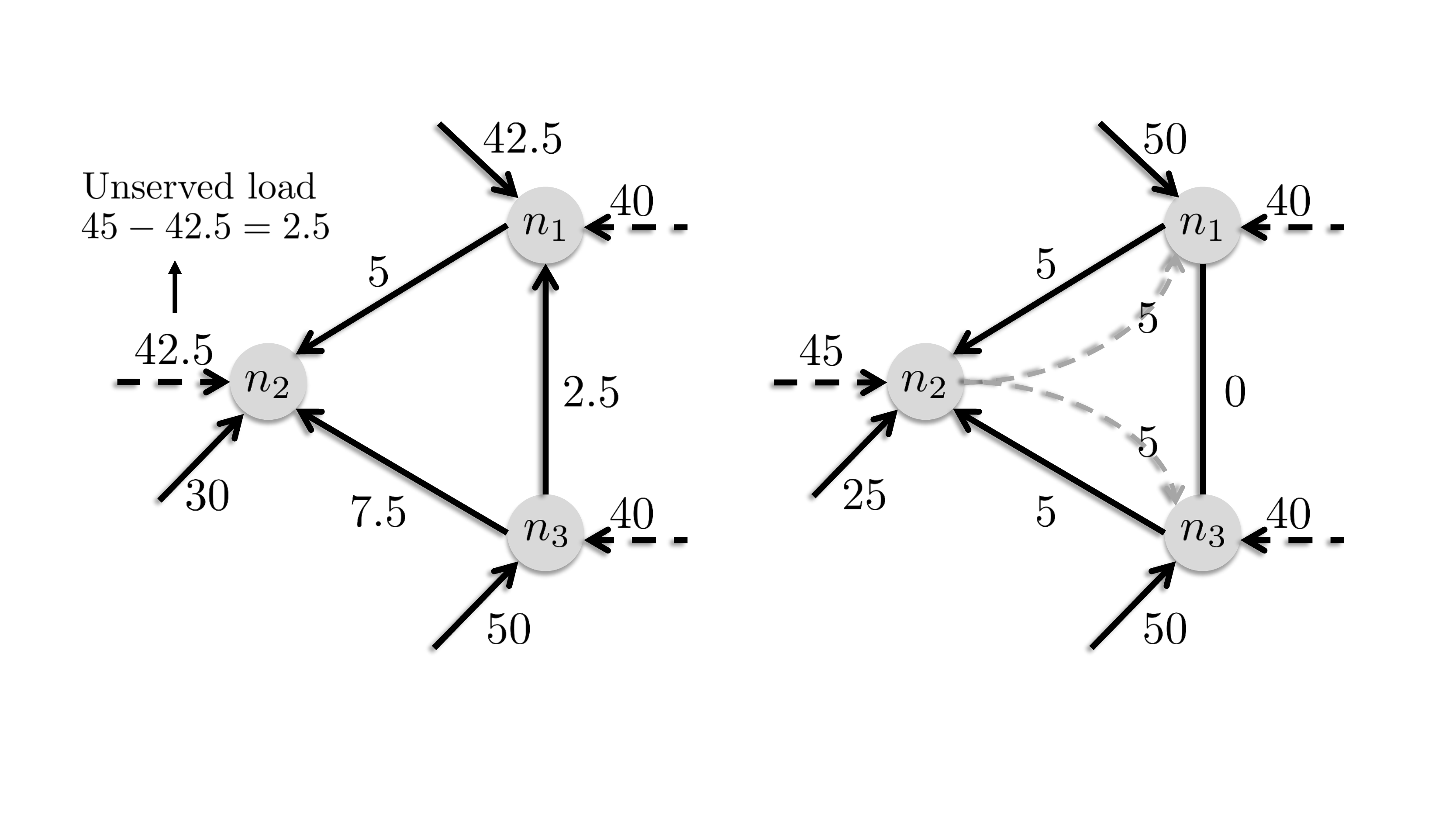}
	\caption{\small The clearing outcome of 3-bus case study scenarios 1 (left) and 4 (right). The black dashed lines denote loads, black solid lines supplies or power transmission (if between two nodes). Grey dashed curves denote virtual links.}
	\label{fig:3bus}
\end{figure}

The scenarios and results are summarized in Table \ref{table:3bus_results}. Part of the results is illustrated in \ref{fig:3bus}. Here, $\bar{\delta}$ and $\alpha_v^\delta$ denote the maximum shifting capacity and bidding cost for all virtual links. Scenario 1 corresponds to the case in which no spatial flexibility is available from data centers. Scenarios 2 to 5 gradually increase shifting capacities. In scenarios 6 and 7 we use the same shift capacities as scenario 4 but we decrease the shifting cost to show interplay with transmission costs. 

The results show that the social welfare increases as data centers offer more capacity to shift loads, due to an increase in the amount of load served and the more efficient use of generation. Interestingly, in scenarios 3 and 4, transmission flows decrease as more load is served (even though the bid cost of physical transmission is lower than that of virtual links). This is because physical transmission line flows are limited by the line constraints and the Kirchoff laws, but can be overcome using virtual links. Scenario 6 shows that, when the shift cost is lower than the transmission cost, no physical transmission is used at all and instead loads are reconfigured using virtual shifts to obtain the highest social welfare. This illustrates how virtual shifting can provide an economic alternative to transmission and that this can create incentives for the provision of load flexibility by data centers. 

Table \ref{table:3bus_results} also shows that the nodal prices become more homogeneous across the network as the system gains more spatial shifting flexibility. For scenarios 1 to 4, the range of the prices shrinks as the data centers provide more shifting capacity. We also observe that the nodal price distribution exhibits convergence as flexibility increases. Specifically, scenarios 4 and 5 show that, once the prices converge, additional shifting flexibility does not fully homogenize prices. This is related to another observation that, at the limiting solution, the price difference between nodes is the cost of spatial shift. This result can be established from duality and is analogous to the well-known result that nodal price differences are bounded by transmission costs \cite{sampat2019coordinated}. 
The virtual shift cost can be understood as the minimum incentive the market should provide 
to compensate for the cost of the spatial shift. We also observe that in scenario 7, with zero shifting costs, the nodal prices can be made fully homogeneous as shifting capacity is increased.

\subsection{Temporal Flexibility in a One-Bus Network}

\begin{table*}[!h]
\centering
\caption{\small Results for one-bus network with temporal shifting flexibility. }
\begin{tabular}{|c|c|c c c c c|} 
 \hline
 Scenario & $\bar{\delta}$ (MWh) & $\phi$ (\$) & $\pi$ (\$/MWh) & $d$ (MWh) & $p$ (MWh) & $\delta$ (MWh)\\ 
 \hline
 1 & [0,0,0,0] & 5200 & [30,-30,40,15,20] & [40,20,40,40,40] & [40,20,40,40,40] & [0,0,0,0] \\ 
 2 & [10,0,0,0] & 5770 & [30,0,40,15,20] & [60,20,50,40,40] & [50,30,50,40,40] & [10,0,0,0] \\
 3 & [21,0,0,0] & 5840 & [23,20,40,15,20] & [70,20,50,40,40] & [50,40,50,40,40] & [20,0,0,0] \\
 4 & [21,20,0,0] & 5840 & [23,20,40,15,20] & [70,20,50,40,40] & [50,40,50,40,40] & [20,0,0,0] \\
 5 & [21,0,21,0] & 6060 & [23,20,40,37,20] & [70,20,60,40,40] & [50,40,50,50,40] & [20,0,10,0] \\
 6 & [21,0,21,10] & 6200 & [23,20,26,23,20] & [70,20,70,40,40] & [50,40,50,50,50] & [20,0,20,10] \\
 7 & [100,100,100,100] & 6200 & [23,20,26,23,20] & [70,20,70,40,40] & [50,40,50,50,50] & [20,0,20,10] \\
 \hline
\end{tabular}
\label{table:5times_results}
\end{table*}

We now consider a one-node network with one supplier and one data center that offers temporal flexibility over a time horizon of five points. The virtual links are defined to create a storage-like system: at each time node the data center receives loads that are shifted from the previous time and any unserved load is carried over. As boundary conditions we have that, at $t=t_1$, the data center receives no shifted loads while, at $t=t_5$, the data center cannot shift any load to the next time. Within the time window, the load capacity and bid costs of loads and supplies change with time. The system thus have $T=5$ time nodes and $T-1$ virtual links $\mathcal{V}:=\{(1,2), (2,3), (3,4), (4,5)\}$. The generation capacities are set to $\bar{p}=\{50,50,50,50,50\}$, load capacities to $\bar{d}=\{70,20,70,40,40\}$, generation bidding costs to $\alpha^g=\{10,20,10,15,20\}$, and load bidding prices to $\alpha^d=\{30,60,40,50,45\}$. We fix the bidding cost for virtual links as $\alpha^\delta=\{3,3,3,3\}$. For ramp limits, we set $\bar{\Delta p} = 20$. 

\begin{figure}[!ht]
	\centering
	\vspace{-10pt}
	\includegraphics[width=3.7in, trim={1cm 4cm 0 3.5cm},clip]{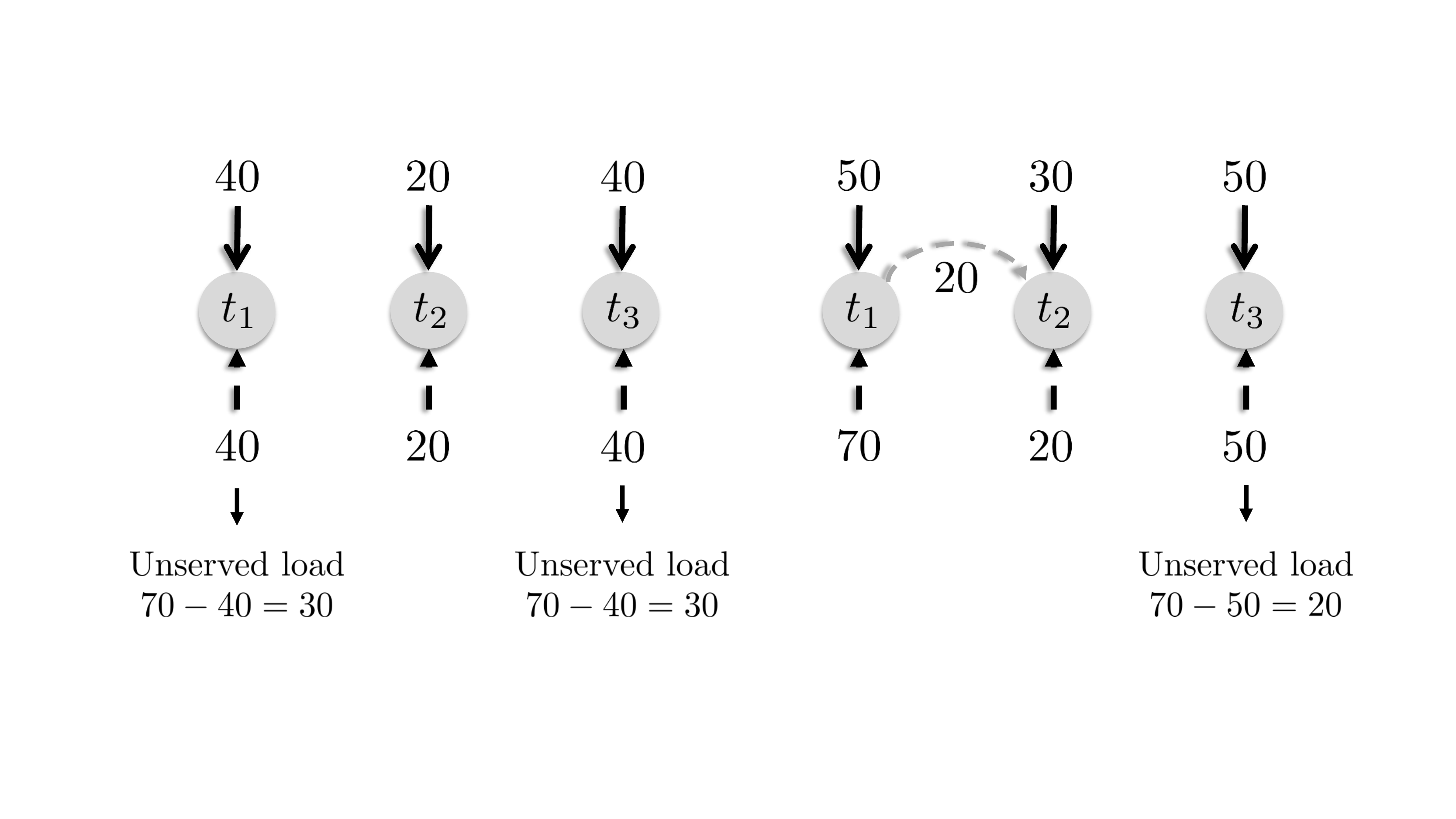}
	\caption{\small The clearing outcome of 5-time case study scenarios 1 (left) and 3 (right). Only the first 3 time points are shown. The black dashed lines denote loads, black solid lines supplies or power transmission (if between two nodes). Grey dashed curves denote virtual links.}
	\label{fig:5times}
\end{figure}

The problem setup follows the sketch in Figure \ref{fig:5times}. We use seven scenarios of different temporal shift capacities to demonstrate important properties, presented in Table \ref{table:5times_results}. The results are analogous to those observed in the spatial shifting case (this highlights how virtual links facilitate treating space-time dimensions in a unified framework). Specifically, the social welfare and the total amount of delivered loads increase with increasing shift capacity. The price variance over the time nodes becomes narrower as shifting capacity is offered. In the limit of high shifting capacity, prices converge and the differences between nodes are bounded by the shifting cost. This illustrates how properties induced by spatial and temporal virtual links are analogous. We note that scenario 1 has a negative LMP caused by the ramping limit, which is relieved in later scenarios by virtual links. Another interesting observation (also shown in figure \ref{fig:5times}) is that for scenarios 1 to 3, even if only a virtual link between $t_1$ and $t_2$ is added, the amount of load cleared at $t_3$ increases. These show how temporal flexibility is able to relieve ramping constraints. A district property of the temporal shifts used here, however, is that their effect on the price gaps is unidirectional. In particular, due to the fact that loads can only be shifted forward in time, temporal shifts of loads can take advantage of a potentially lower price in the future for more revenue, but not vice versa. This property is illustrated in scenarios 3 and 4, where additional flexibility in link $v=(2,3)$ does not change the solution since the price at node 2 is higher than that at node 3. 

\subsection{Space-Time Flexibility in IEEE 30-Bus Network}

\begin{figure}
    \centering
    \includegraphics[width=2.2in,trim={0 0 0 0},clip]{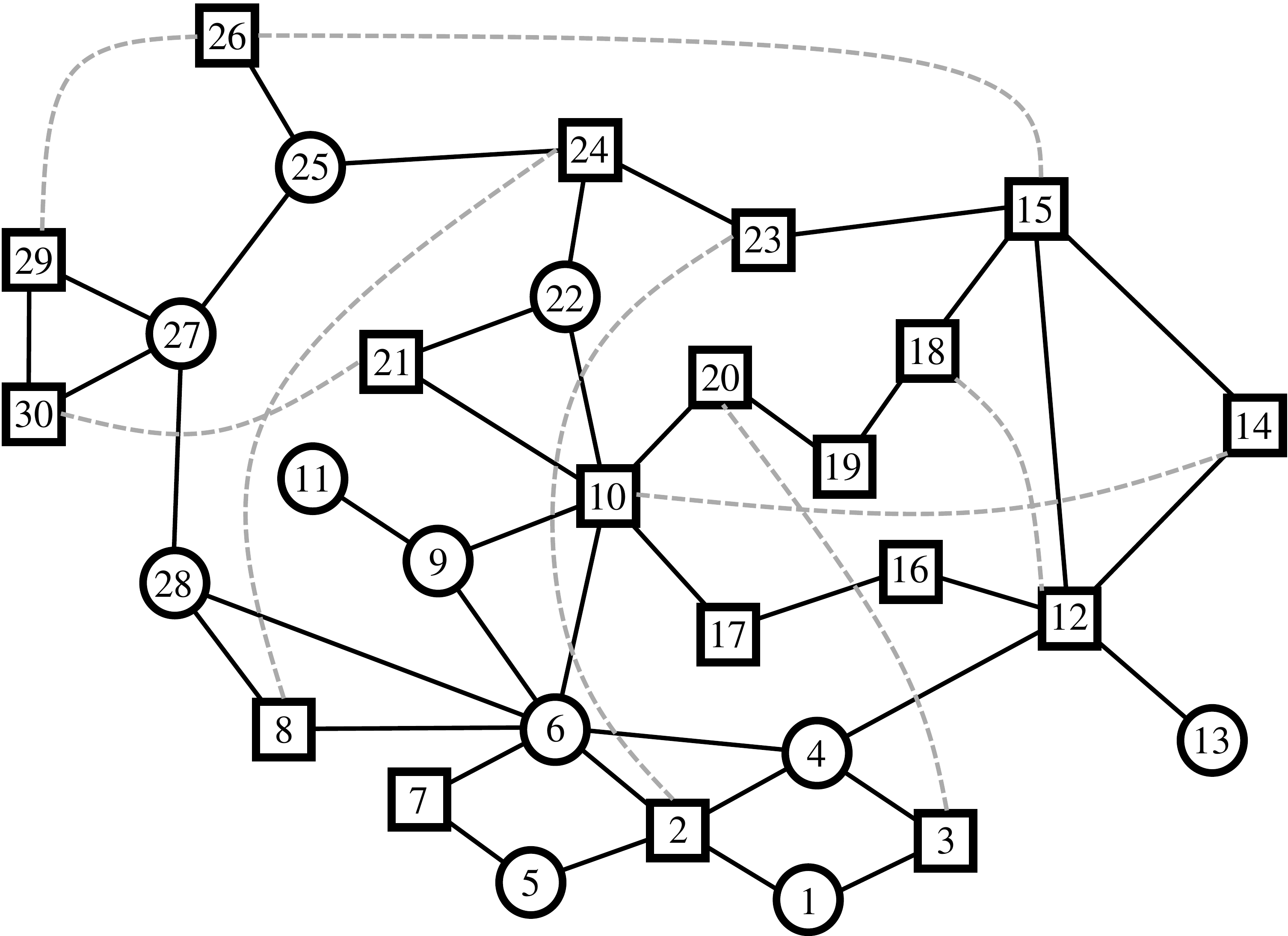}
    \caption{\small Schematic of IEEE 30-bus system showing transmission links (solid) and spatial virtual links (dotted). Nods are either attached to loads (square) or not (circle). Only a fraction of spatial virtual links are shown for clarity.}
    \label{fig:ieee_30_bus}
\end{figure}

We now consider the provision of space-time flexibility over the IEEE 30-bus power distribution network. For simplicity, we consider a time window of three nodes. Here, we impose an upper bound for the spatial and temporal shits of at most 10 units of power. A schematic of the network with spatial virtual links is shown in \ref{fig:ieee_30_bus}. The optimization problem is solved with and without the data center flexibility.

\begin{figure*}[!ht]
    \centering
    \begin{tabular}[t]{c c c c c}
        \begin{tabular}{c}
        \smallskip
             \begin{subfigure}[b]{0.27\textwidth}
                 \centering
                 \includegraphics[width=0.95\textwidth]{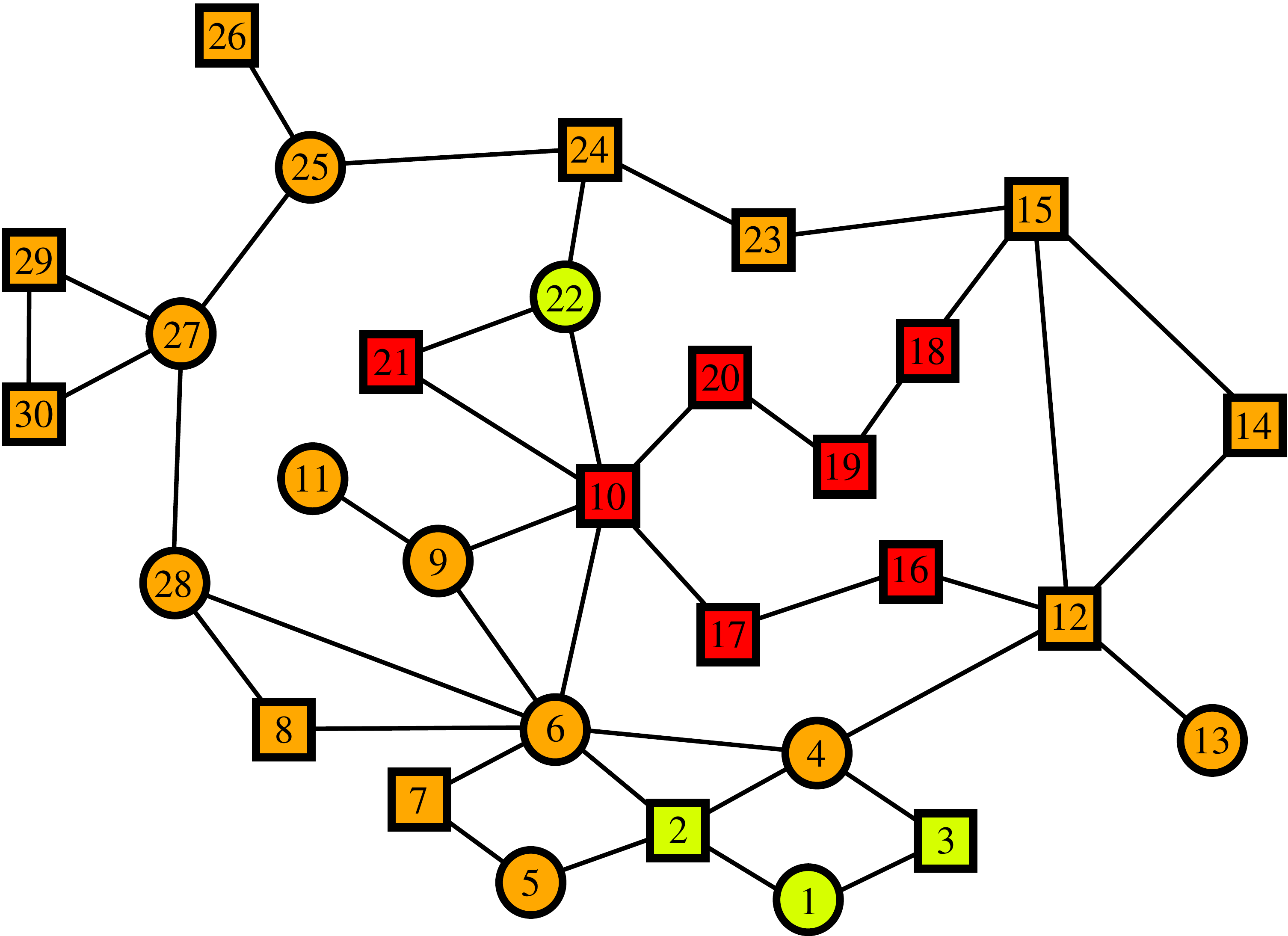}
                 \caption{\small $t=1$, $\sigma = 9.67 \$$, $\text{MAD} = 5.67\$$}
                 \label{fig:t1_no_shift}
             \end{subfigure}
             \\
             \begin{subfigure}[b]{0.27\textwidth}
                 \centering
                 \includegraphics[width=0.95\textwidth]{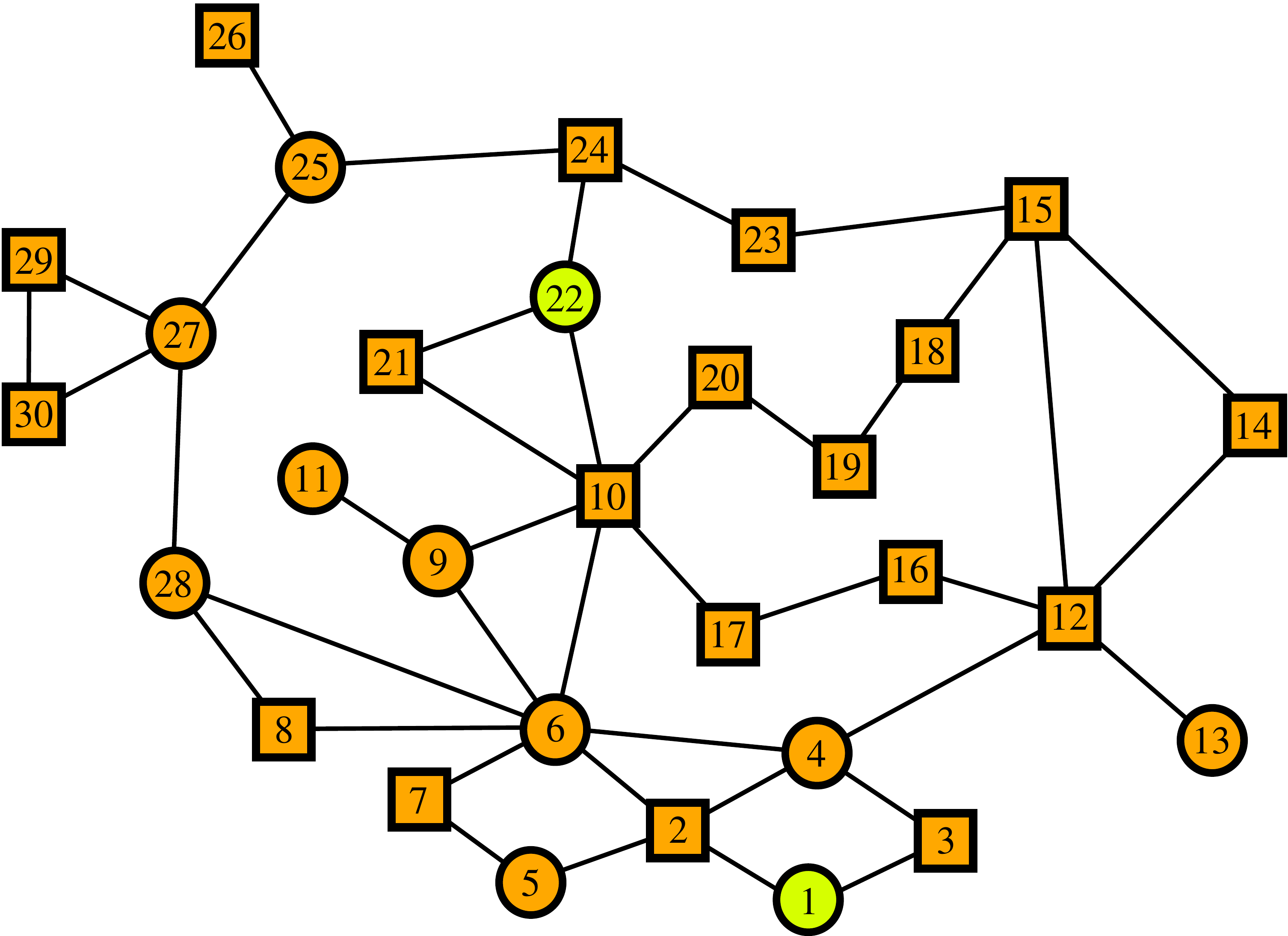}
                 \caption{\small $t=1$, $\sigma = 4.43\$$, $\text{MAD} = 0.82\$$}
                 \label{fig:t1_shift}
             \end{subfigure}
        \end{tabular}
        \begin{tabular}{c}
        \smallskip
             \begin{subfigure}[b]{0.27\textwidth}
                 \centering
                 \includegraphics[width=0.95\textwidth]{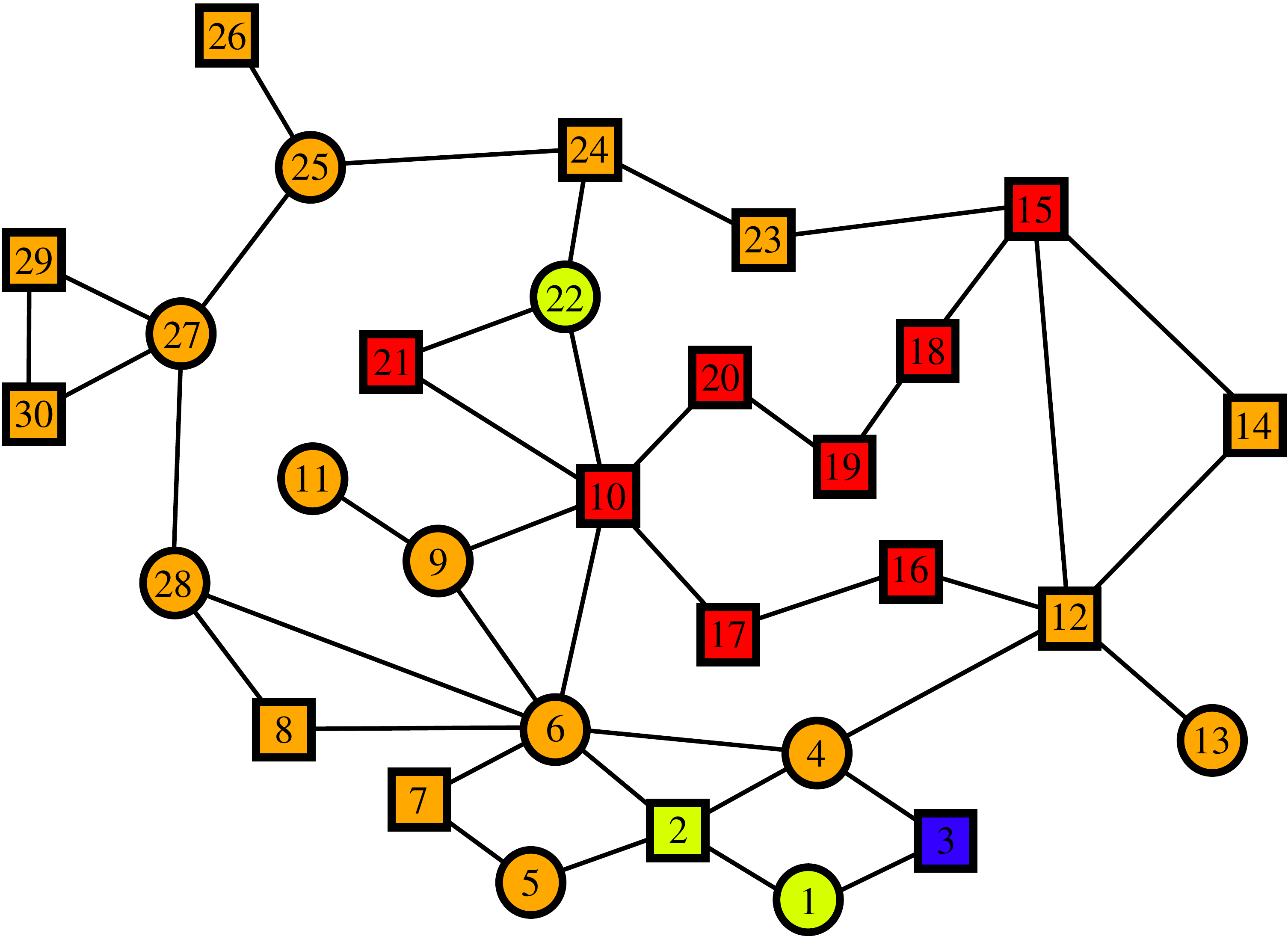}
                 \caption{\small $t=2$, $\sigma = 14.6 \$$, $ \text{MAD}= 7.83\$ $}
                 \label{fig:t2_no_shift}
             \end{subfigure}
             \\
             \begin{subfigure}[b]{0.27\textwidth}
                 \centering
                 \includegraphics[width=0.95\textwidth]{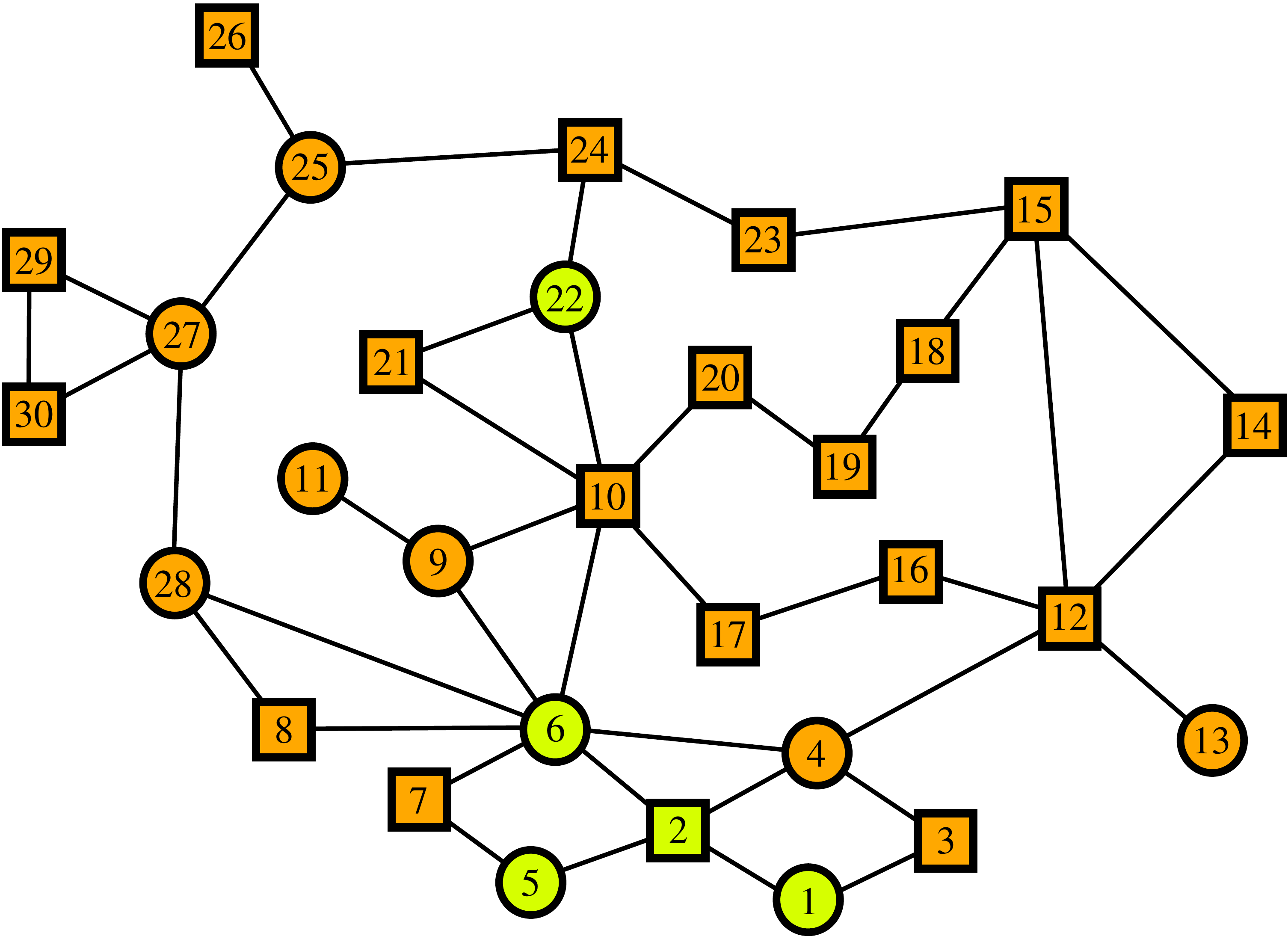}
                 \caption{\small $t=2$, $\sigma = 5.59\$$, $\text{MAD} = 1.73\$$}
                 \label{fig:t2_shift}
             \end{subfigure}
        \end{tabular}
        &
        \begin{tabular}{c}
        \smallskip
             \begin{subfigure}[b]{0.27\textwidth}
                 \centering
                 \includegraphics[width=0.95\textwidth]{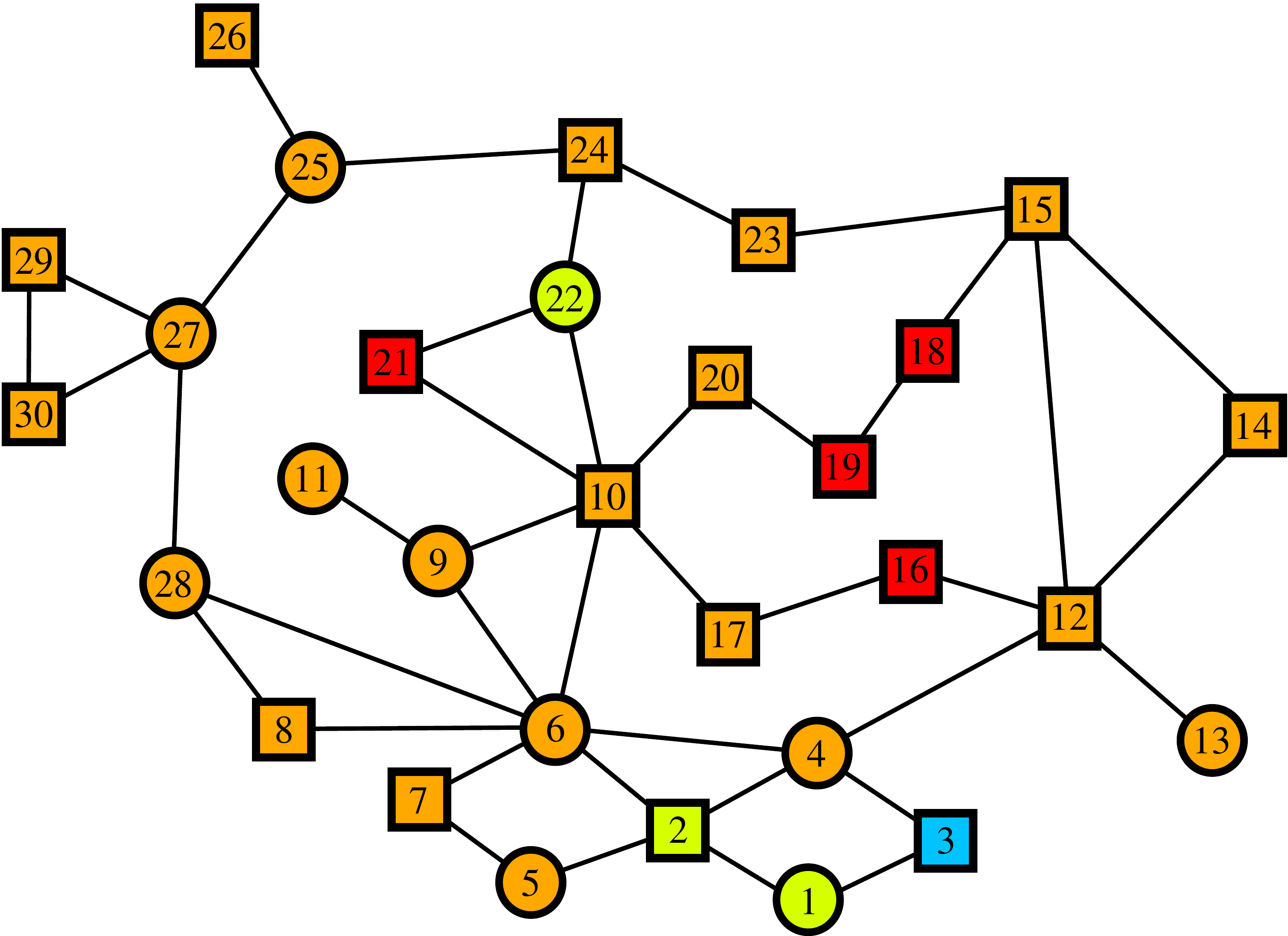}
                 \caption{\small $t=3$, $\sigma = 10.7\$$, $\text{MAD} = 5.12\$$}
                 \label{fig:t3_no_shift}
             \end{subfigure}
             \\
             \begin{subfigure}[b]{0.27\textwidth}
                 \centering
                 \includegraphics[width=0.95\textwidth]{colored_vl_t1.pdf}
                 \caption{\small $t=3$, $\sigma = 4.09\$$, $\text{MAD} = 1.52\$$}
                 \label{fig:t3_shift}
             \end{subfigure}
        \end{tabular}
        &
        \begin{tabular}{c}
        \smallskip
            \begin{subfigure}[b]{0.08\textwidth}
                \centering
                \includegraphics[width=0.7\textwidth]{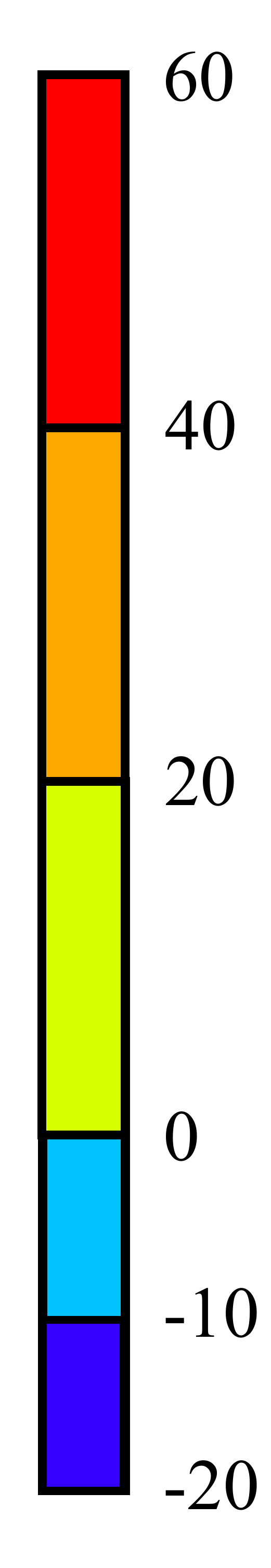}
                \label{fig:color_map1}
            \end{subfigure}
        \end{tabular}
    \end{tabular}
    \caption{\small Space-time prices for IEEE 30-bus system without (top row) and with (bottom row) virtual links. The LMP value of each node is denoted by its color. The variance ($\sigma$) and mean absolute deviation (MAD) values of LMP at each time point are shown in the subtitles.}
    \label{fig:30-bus_results}
\end{figure*}

Figure \ref{fig:30-bus_results} shows the space-time price distributions. For the case without virtual links, the heat maps (Figures \ref{fig:t1_no_shift} to \ref{fig:t3_no_shift}) show a high level of spatial and temporal price volatility. Specifically,  we can observe multiple congestion nodes (nodes with prices greater than or equal to 40) and several nodes with low prices and the price range increases at certain times. This spatial and temporal price volatility is relieved with the addition of virtual shifts. Specifically, as shown in Figures \ref{fig:t1_shift} to \ref{fig:t3_shift}, at each time point the price range shrinks. Moreover, we have found that the social welfare increases by 100\% (from \$5,209 to \$11,217) by incorporating virtual links. 

From Figure \ref{fig:30-bus_results} we observe clustering behavior for the space-time prices. For example, in the case without virtual links, the prices of nodes 10, 17, 20 remain at the same level, while the price of nodes 2 is noticeably lower throughout the entire time window. When virtual links are added the prices become more homogeneous and follow a similar trend. We note that this homogenization behavior is not limited to cluster of nodes that are close to each other geographically. For instance, nodes 21 and 30 have an associated data center and therefore are connected by a spatial virtual link. Due to the added shifting flexibility, they also exhibit the price homogenization, even though they are far apart. This illustrates how virtual links can help overcome geographical barriers associated with transmission network topology. 

In this study we also observed the emergence of negative prices. For the case without data center flexibility, node 3 has negative prices at time 2 and 3. In the cases of negative prices, electricity flows from node 1 to node 3, and from node 3 to node 4. This is caused by transmission constraints. When the system incorporates virtual shifting, space-time price volatility reduces and negative prices disappear. 

\section{Conclusions and Future Work}
\label{sec:conclusion}
\vspace{-0.05in}
We presented the concept of virtual links as a way to capture space-time load shifting flexibility of data centers in market clearing. We show that virtual links are mathematically equivalent to transmission links, which facilitate the analysis and interpretation of the clearing quantities and prices. Moreover, we show that virtual links can be used to capture space time behavior in a systematic manner. Case studies show that virtual shifts lead to higher social welfare, higher load served, and mitigation of space-time price volatility. Our results also highlight that virtual links provide an additional source of revenue for data centers and thus create incentives to provide flexibility. Also, the result is consistent with previous observation that flexibility limits incentives for market manipulation between transmission and energy markets \cite{birge_hortacsu_pavlin_2017}.

\section*{Acknowledgements}
\vspace{-0.05in}
We acknowledge support from the U.S. NSF under award 1832208. The fourth author acknowledges support from the University of Chicago Booth School of Business.

\bibliographystyle{IEEEtran}\bibliography{refs}

\end{document}